# Phase transition in magnetically coupled spins on a ring (SOR) model

Rajib Biswas and Biman Bagchi
Solid State and Structural Chemistry Unit,
Indian Institute of Science, Bangalore – 560012
Email – rajib@sscu.iisc.ernet.in, bbagchi@sscu.iisc.ernet.in

*Abstract*

We have considered a new type of 'XY' model where spins are placed on concentric ring with constant spin density in every ring. The spin executes continuous rotation under a modified Shore-Zwanzig Hamiltonian (J. Chem. Phys. 63, 5445 (1975)). We have performed Monte Carlo simulation using Glauber acceptance criteria. Computations of Binder's cumulant, specific heat and magnetic susceptibility all show presence of a finite temperature order-disorder phase transition in this spin system. The system size dependence of Binder's cumulant suggests the existence of a phase transition with a transition temperature of $T^* = 1.2$. However, we have found no signature of the occurrence of vortex in our SOR model. The absence of hysteresis rules out the possibility of first order phase transition. We have found two "stable" states for $T^* = 0$ phase. The perfectly ordered true ground state is obtained by gradual cooling of the system, while the other is obtained by starting the simulation with a random configuration at $T^* = 0$. This second state has higher energy than the perfectly aligned ground state.

I. Introduction

Variations of original Ising spin Hamiltonian and lattice type are found to be useful in modeling the interacting systems in condensed matter. It is well known that modified statistical model systems such as random Ising and the XY models show completely different types of phase behavior compared to their regular counterparts. The conventional 2D XY model does not possess any magnetization at non-zero temperature[1]. However, the presence of topological defects leads to a quasi-long-range-order to disorder phase transition known as Kosterlitz-Thouless (KT) phase transition[2]. It is also well known in the literature that the modification of XY lattice as well as interaction potential lead to dramatic change in the character of the phase transition[3,4].

In Monte Carlo studies of critical phenomena, renormalization group invariant quantities are the key quantities. In finite size scaling (FSS) these quantities permit us to identify the transition point, to determine the nature of the transition. An example of such quantity is the Binder cumulant[5]:

$$U_N = 1 - \frac{\langle M^4 \rangle}{3\langle M^2 \rangle^2} \tag{1}$$

where $M$ is the total magnetization of the system. While the Binder cumulant is a standard parameter in the study of second-order transitions, only a few examples are known in the case of KT transition[3,6]. This is because very little is known about the behavior of the Binder cumulant in the vicinity of the KT transition.

In this work we introduced a new variant of spin model, spin on a ring or, SOR model, which is described in **Fig. 1**. This model has been earlier employed to mimic a dipolar liquid confined within an otherwise empty spherical object, such as water molecules confined within a reverse micelle[7]. In the latter system, the surface can induce the molecules to organize in layers and drive the system towards an orientational transition. The comparison with the usual XY model sheds light on the role of the geometry and coupling parameter on the phase transition. Monte Carlo computations of Binder's cumulant[5], specific heat and magnetic susceptibility show the presence of finite temperature phase transition in this newly developed SOR model.

We arrange the rest of the paper as follows: Section – II describes the model and simulation details used in the present work. The definition of the observables and discussion about the results are in section – III.

## II. Model and simulation details

Our SOR model consists of collection of **s**pins at a constant number density, placed on several concentric rings[7] of unit inter ring spacing. All spins execute continuous rotation in 'XY' plane using modified Shore-Zwanzig[8] Hamiltonian

$$H = -\sum_{ij} J_{ij}(r_{ij}) \cos(\theta_i - \theta_j), \tag{2}$$

where $J_{ij}(r_{ij}) = J_0 \exp(-r_{ij}/\lambda)$, is the coupling constant between *i*-th and *j*-th spin separated by $r_{ij}$ and $\theta_i$ is the spin angle defined with respect to the *x*-axis. The total number of spins in the *i*-th ring, $N_i = \delta_{i1} + 4R_i$, where $R_i$ is the radius of *i*-th ring. On each ring the spins are equally spaced with an angle, $2\pi/N_i$. There is only one ring at the centre.

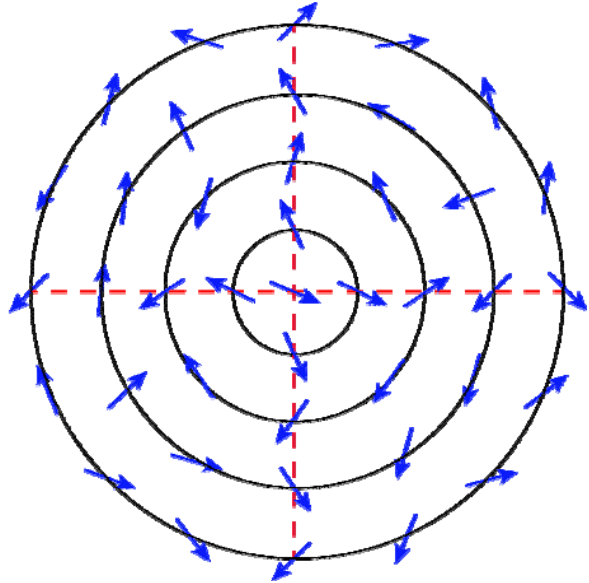

**Fig. 1**: **Schematic representation of the spin on a ring (SOR) model. It consists of several concentric rings of identical spin density. The spatial direction of every spin is represented by a blue arrow. Note that there is no periodic boundary condition in this SOR model.**

We have performed single spin flip Monte Carlo simulation using Glauber's[9] acceptance criteria. We have carried out Monte Carlo simulations for 10, 20 and 30 ring systems, using $J_0 = 1.0$, and $\lambda = 1.0$ for a temperature range of 0.0-3.0 with increment of 0.1. The MC simulations have been carried out initially for an equilibration run of $3 \times 10^5$ MC steps and the subsequently $2 \times 10^5$ MC steps for data collection. We have also carried out a set of MC simulations with the modified Hamiltonian (given by Eq.(2)) for 16×16, 24×24 and 32×32 square lattices. Results and analyses are presented below.

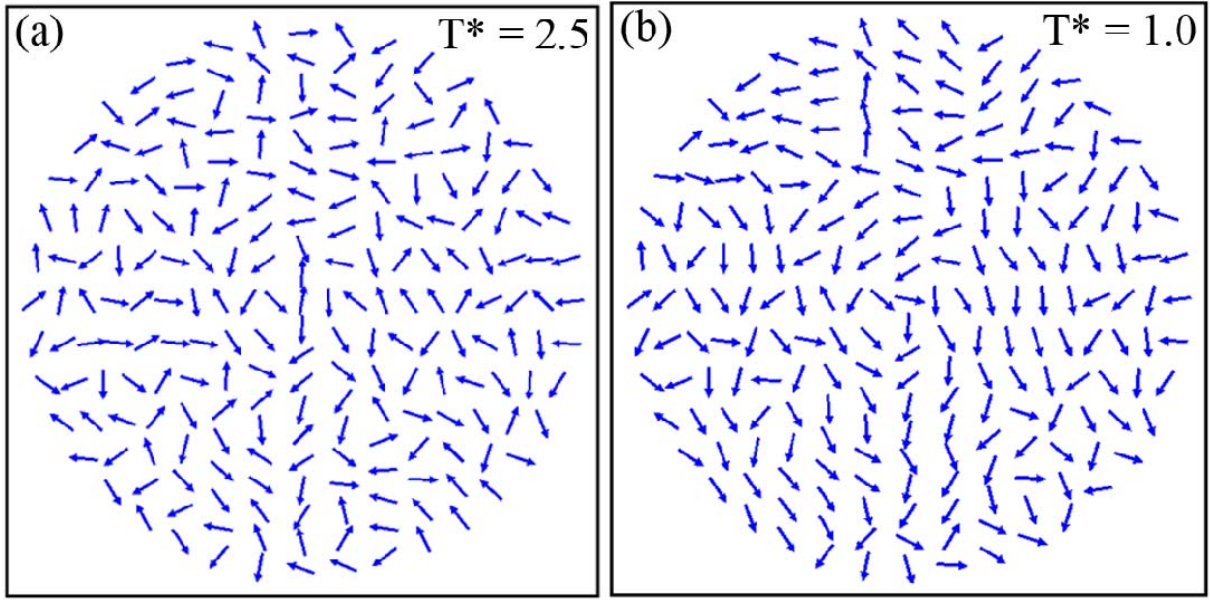

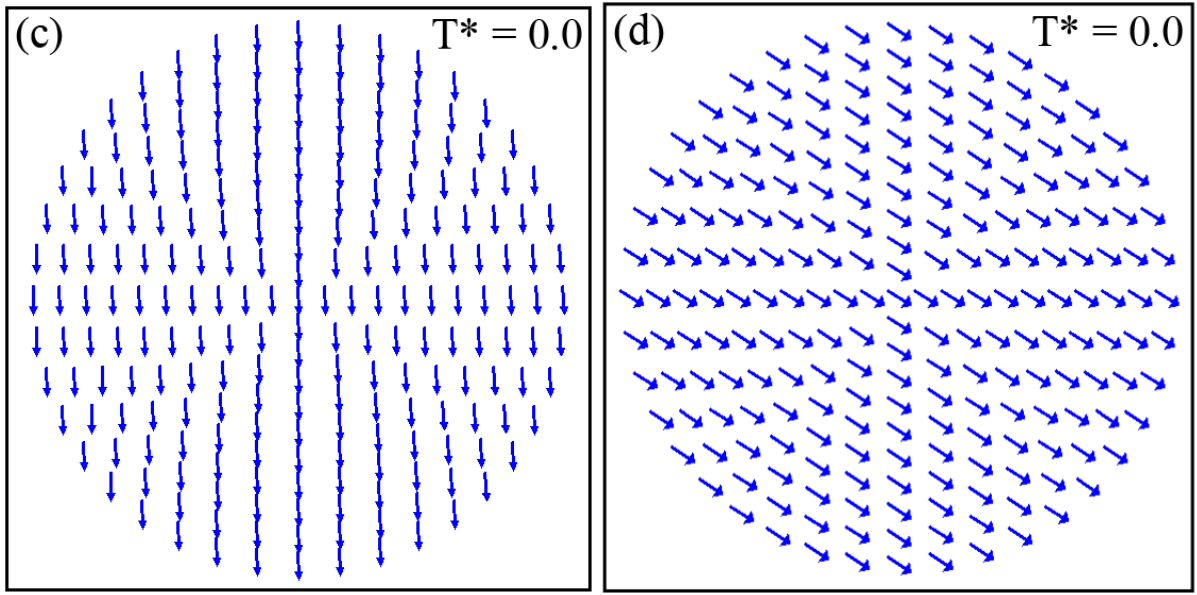

**Fig. 2:** Snapshot of 10 ring system with $J_0 = 1.0$, $\lambda = 1.0$, and at (a) $T^* = 2.5$, (b) $T^* = 1.0$, (c) and (d) $T^* = 0.0$. The state shown in (c) is obtained by gradual cooling from high temperature while state shown in (d) is obtained by starting the simulation with random configuration at $T^* = 0.0$. Note that state shown in (d) is of lower energy than that of (c).

### III. Results and discussions

We have already mentioned that Binder's cumulant (BC) permits to locate the critical point (if present) and critical exponents from simulation of finite sized systems. For an Ising model with zero external magnetic field, BC is given by Eq. (1). For SOR model, magnetization $M$ is given by

$$M = \frac{1}{N}\sum_{j=1}^{N} e^{i\theta_j} \qquad (3)$$

where $\theta_j$ is the orientation of the $j$-th spin. We have computed Binder's cumulant for SOR as well as XY square lattice using modified Hamiltonian (Eq. (2)). As it was already reported in literature that 2D square lattice shows merging of $U_N$ for different $N$'s in the whole low-temperature phase with quasi-long-range order[3], while the SOR model, apart from the similar kinds of merging of $U_N$, also exhibits a unique crossing at a well-defined critical temperature leading to the separation of truly ordered phase and disordered phase **(Fig. 3b)**. From **Fig. 3,** it is clear that there exist a phase transition at a temperature T* = 1.2. This has been further supported by calculations of magnetic susceptibility as well as specific heat. The specific heat $C_V$ and susceptibility $\chi$ have been calculated by using the following fluctuation formulae

$$C_V = \frac{1}{N} \frac{\langle E^2 \rangle - \langle E \rangle^2}{k_B T^2} \tag{4}$$

$$\chi = \frac{\langle M^2 \rangle - \langle M \rangle^2}{k_B T} \tag{5}$$

where *M* is given by Eq. (3). Below we show the results in **Figure 3.**

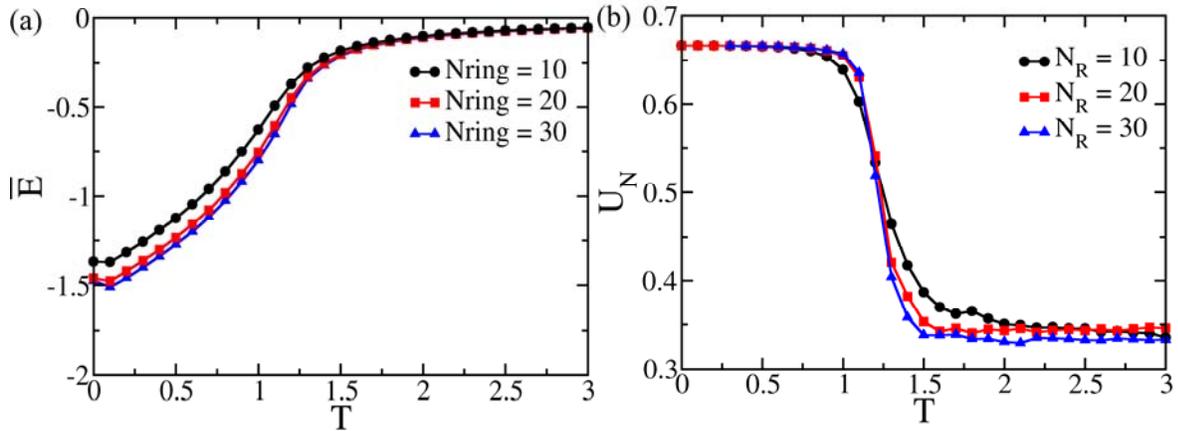

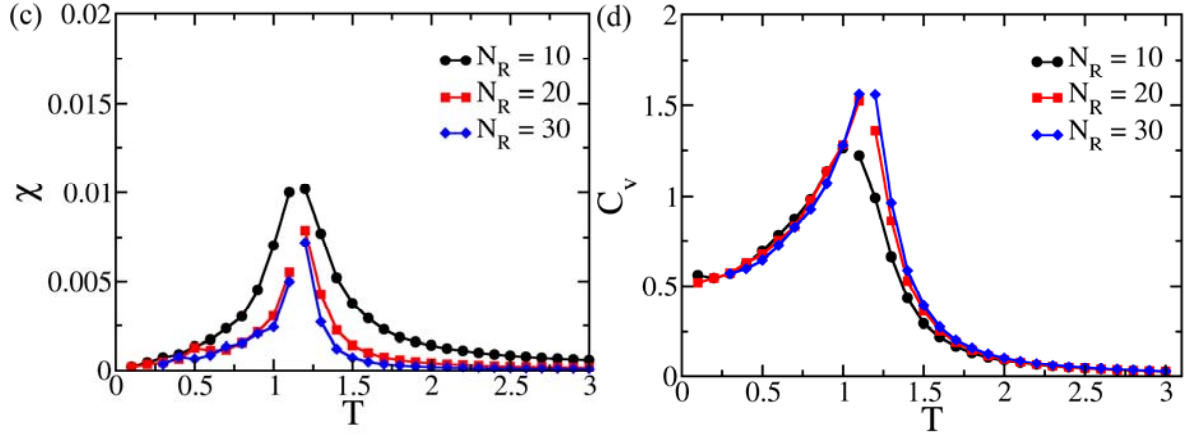

**Fig. 3:** Monte Carlo simulation results of the spin on a ring (SOR) model for $J_0=1.0$ and $\lambda=1.0$: (a) Average energy per spin vs. temperature, (b) Binder's cumulant vs. temperature, (c) magnetic susceptibility vs. temperature and (d) specific heat vs. temperature. The unique crossing point of Binder's cumulant suggest a well-defined critical temperature that separates the truly ordered phase from the disordered phase.

For comparisons we show in **Fig. 4** the analogous properties calculated for the X-Y model on a square lattice with our modified Hamiltonian. Note that there are notable similarities, although there are minor differences in details.

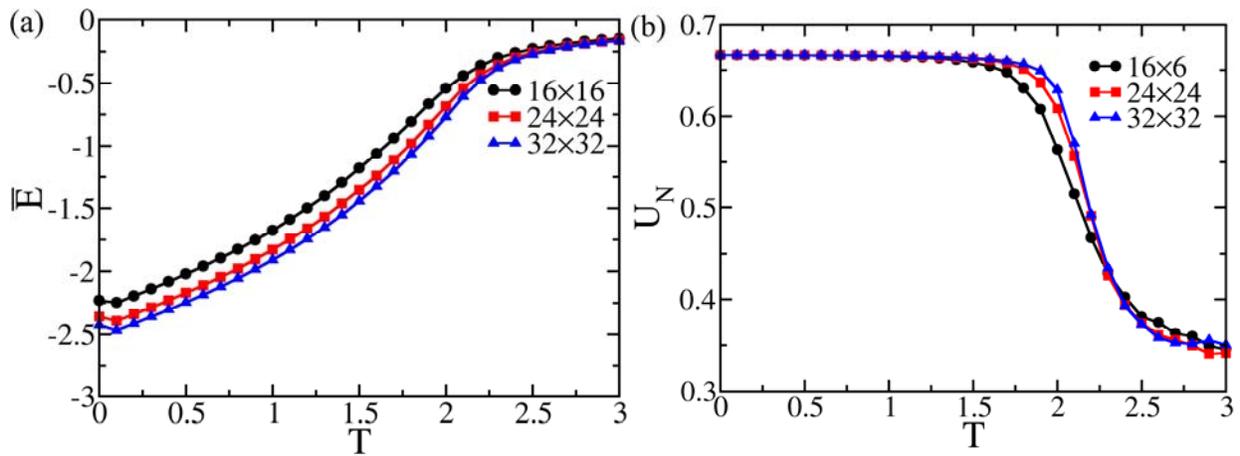

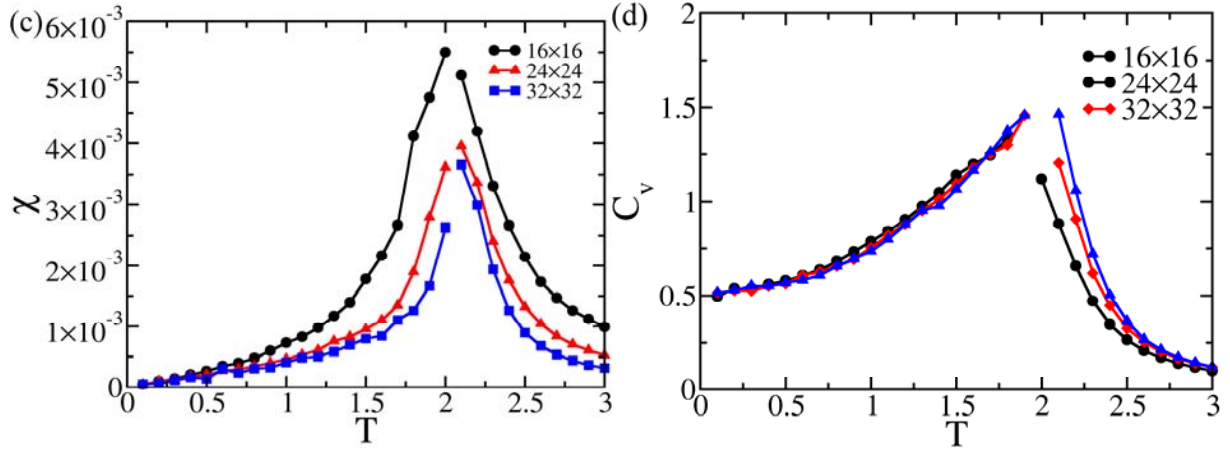

**Fig. 4:** Results of XY model with our modified Hamiltonian using $J_0=1.0$ and $\lambda=1.0$: (a) Average energy per spin vs. temperature, (b) Binder's cumulant vs. temperature, (c) magnetic susceptibility vs. temperature and (d) specific heat vs. temperature. The unique crossing point of Binder's cumulant suggest a well-defined critical temperature by splitting the truly ordered phase and disordered phase.

It is clear from **Fig. 3** and **Fig. 4** that the appearance of a unique crossing point in Binder's cumulant is not due to the geometry of SOR model. This is due to the modified Hamiltonian. The distance dependence coupling constant facilitates formation of long range order in the system. As a result the system tends to show nonzero magnetization at finite temperature. One can tune the value of $J_0$, and can easily observe a critical value of $J_0$ for which there can be a true phase transition rather than KT type transition.

We have also studied the effects of coupling strength on the phase transition behavior in the SOR model. If we increase the $J_0$ value then system exhibits nonzero magnetization up to higher temperature, and the transition temperature shifts towards high value as depicted in **Fig. 5**.

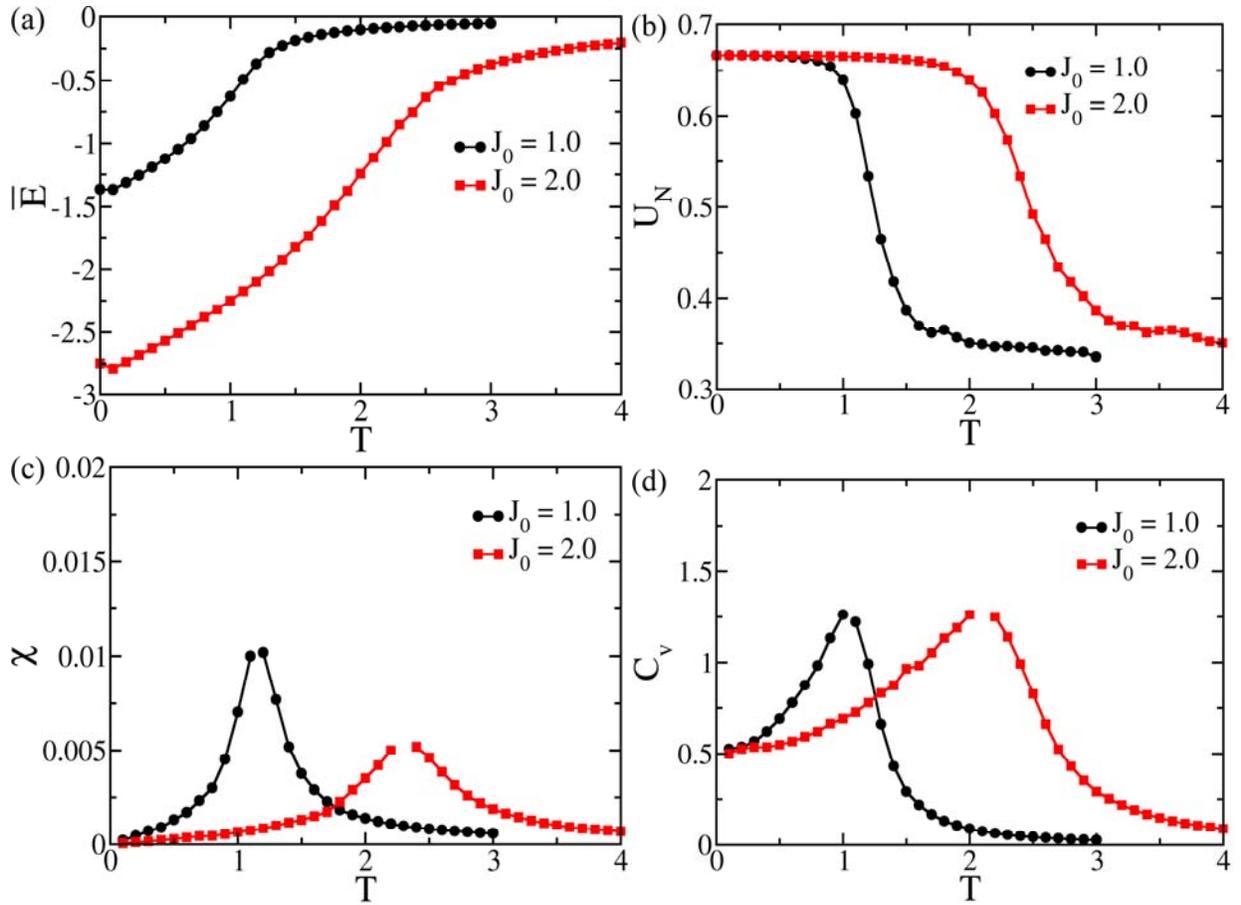

**Fig. 5: Effects of the strength of coupling constant $J_0$ of SOR model: (a) Average energy per spin vs. temperature, (b) Binder's cumulant vs. temperature, (c) magnetic susceptibility vs. temperature and (d) specific heat vs. temperature.**

We have observed no hysteresis in the average energy of the system during a cooling-heating cycle (as shown in **Fig. 6**). The absence of hysteresis clearly rules out the presence of a first order phase transition.

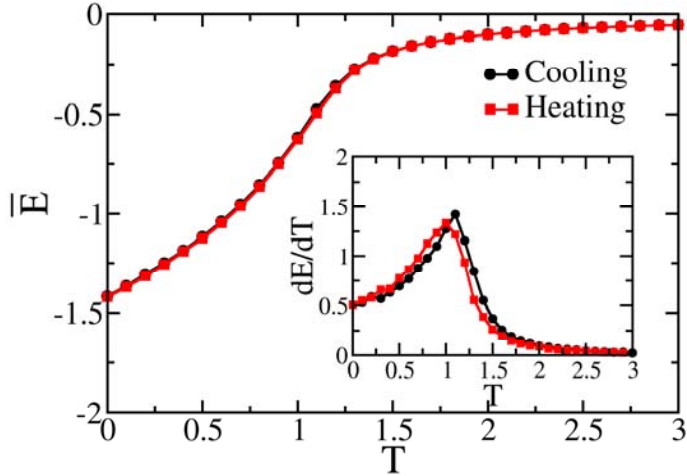

**Fig. 6:** Cooling-Heating cycle for $N_{ring}$ = 10 system. In inset the first derivative of average energy versus temperature is plotted.

## IV.   Conclusion

The model "spin on a ring" was introduced to describe orientational ordering of an interacting system confined to a spherical geometry. This model is substantially different from a square lattice because the neighbors in this case are placed at different distances. While many of the features observed in the SOR model are similar to the ones observed in the square lattice model, several things are also different. The most important difference is the absence of the vortices in the present system. This makes the transition from the disordered to ordered phase a bit stronger than that observed in the well-known X-Y model on the square lattice.

## V.  ACKNOWLEDGEMENT

We thank DST and CSIR (India) for partial support of the work reported here. We also thank JC Bose Fellowship (DST) for support.